\documentclass[oneside,12pt,dvips]{article}

\usepackage{amsfonts,amssymb,epsfig}

\begin{document}

\section*{
  \begin{center} Measurement of CP violation phase $\phi_s$ and \linebreak
                 charge asymmetries in $B^0_{(s)}$ decays at D\O
  \end{center}
}

\begin{center} \bf L. Sonnenschein \rm \\[1.0ex]
  CERN and LPNHE Paris, Universit\'es Paris VI, VII \\[3ex]
  (for the D\O\ collaboration at the CKM 2008 workshop)
\end{center}

\vspace*{2ex}

\begin{abstract}

The CP violation phase $\phi_s$ and charge asymmetries in $B^0_s$ decays have been measured
by the D\O\ experiment in Run~II of the Fermilab Tevatron Collider where proton
anti-proton collisions take place at a center of mass energy of $\sqrt{s}=1.96$~TeV.
The measurements are based on integrated luminosities between 1.0~fb$^{-1}$ and 2.8~fb$^{-1}$.
A $1.8\sigma$ deviation from the Standard Model is observed in the measurement of the
CP violation phase $\phi_s= -0.57^{+0.24}_{-0.30}$~(stat) $^{+0.07}_{-0.02}$~(syst) rad
in the decay channel $B^0_s\rightarrow J/\!\psi\phi$. A combination with CDF results yields 
a $2.2\sigma$ deviation from the Standard Model.
\end{abstract}

\section*{Introduction}

One of the ultimate challenges in elementary particle physics is the understanding of all 
possible sources of violation of CP symmetry. Within the Standard Model (SM) of particle physics,
CP symmetry is violated through the CKM mechanism~\cite{c1}.
The level of CP violation in the SM is too small to produce the observed baryon number density
in the universe~\cite{h1}.
New phenomena beyond the SM may alter the CP violating mixing phase $\phi_s$ and subsequently 
the decay width difference $\Delta\Gamma_s = \Delta\Gamma_s^{SM}\times |\cos\phi_s|$.
The mixing parameters $\Delta\Gamma_s$ and $\phi_s$ are measured in the decay mode
$B^0_s\rightarrow J/\!\psi\phi$.
Further constraints on these mixing parameters are obtained by measurements of charge asymmetries 
due to $B^0_{(s)}$ oscillations.

\section*{Dimuon charge asymmetry}
The dimuon charge asymmetry $A$ has been measured~\cite{p1} with the D\O\ detector, making use of an 
integrated luminosity of 1.0~fb$^{-1}$. Assuming that the asymmetry $A$ is due to
$B^0 \leftrightarrow \bar{B}^0$ mixing and decay the corresponding CP violation parameter
is being measured to
\begin{equation}
 \frac{{\cal{R}}(\epsilon_{B^0})}{1+|\epsilon_{B^0}|^2}    
 = \frac{A_{B^0}}{4}
 = -0.0023\pm0.0011~(\mbox{stat}) \pm 0.0008~(\mbox{syst}) .
\end{equation} 
Relaxing the assumption in allowing for CP violation contributions from $B^0_s$ systems
yields
\begin{equation}
 A_{B^0}+0.72 A_{B^0_s} = -0.0092\pm0.0044~(\mbox{stat}) \pm0.0032~(\mbox{syst}) .
\end{equation}

\section*{CP violation search in semileptonic $B_s$ decays}
A search for CP violation in semileptonic $B^0_s$ decays has been performed with a data sample 
corresponding to an integrated luminosity of 2.8~fb$^{-1}$. The flavor of the $B^0_s$ meson in 
the final state has been determined by means of the muon charge in the decay 
$B^0_s\rightarrow D^-_s\mu^+\nu X$ with $D^-_s\rightarrow\phi\pi^-$ and $\phi\rightarrow K^+K^-$.
A combined tagging method has been used to determine the initial state flavor. A time-dependent
fit to $B^0_s$ candidate distributions yields the CP violation parameter
\begin{equation}
  a_{sl}^s = -0.0024\pm0.0117~(\mbox{stat}) ^{+0.0015}_{-0.0024}~(\mbox{syst}) .
\end{equation}

\section*{Charge asymmetry in semileptonic $B^0_s$ decays}
This first direct measurement~\cite{p2} of the time integrated flavor untagged charge asymmetry in
semileptonic $B^0_s$ decays. $A_{SL}^{s,\mbox{\scriptsize unt.}}$ has been obtained from a data sample
corresponding to an integrated luminosity of 1.3~fb$^{-1}$ in comparing the decay rate
$B^0_s\rightarrow \mu^+ D^-_s \nu X$, $D^-_s\rightarrow \phi\pi^-$, $\phi\rightarrow K^+K^-$ with its charge conjugated
decay rate. The asymmetry amounts to
\begin{equation}
  A^{s,\mbox{\scriptsize unt.}}_{SL} = [1.23\pm0.97~(\mbox{stat}) \pm0.17~(\mbox{syst})] \times 10^{-2} , 
\end{equation}
assuming that $\Delta m_s / \overline{\Gamma_s}\gg 1$. The result can be further related to the
CP-violating phase in $B^0_s$ mixing via
\begin{equation}
  \frac{\Delta\Gamma_s}{\Delta m_s}\tan\phi_s = [2.45\pm1.93~(\mbox{stat}) \pm0.35~(\mbox{syst})] \times 10^{-2} .
\end{equation}

\section*{$B^0_s$ mixing parameters in the decay $B^0_s\rightarrow J/\!\psi\phi$}

\begin{figure}[t]
\includegraphics[width=12.0cm]{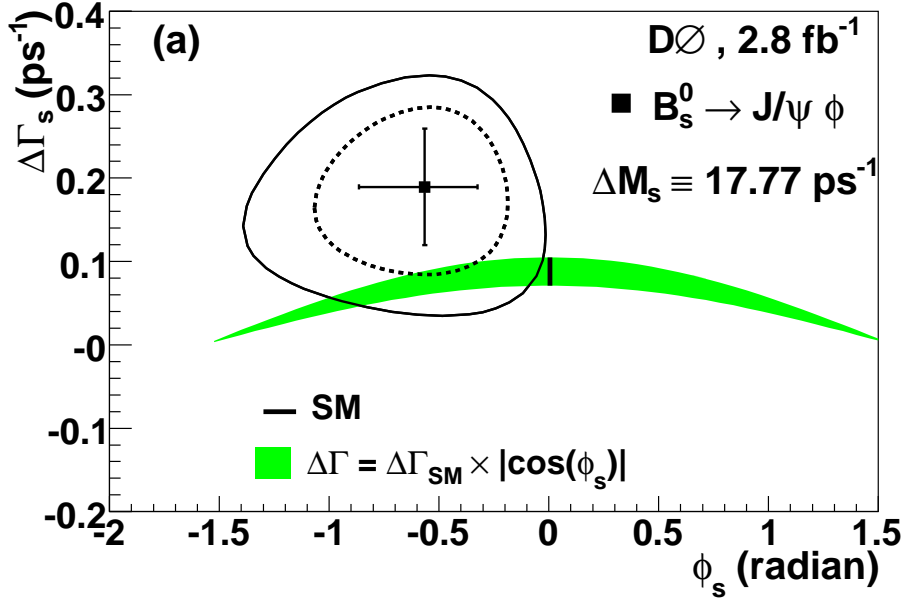}
\vspace*{-1ex}
\caption{\label{phi_s_contour}
The $1\sigma$ and $2\sigma$ 2D contours in the plane ($\Delta\Gamma_s, \phi_s$) for the fit
to $B^0_s\rightarrow J/\!\psi\phi$ (dashed and full line) and the best fit (square) with its 1D 
$1\sigma$ uncertainties (cross), both with constraints on strong phases. 
The SM prediction is also indicated.}
\end{figure}

\begin{figure}[p]
\hspace*{-3ex}
\includegraphics[width=7.15cm]{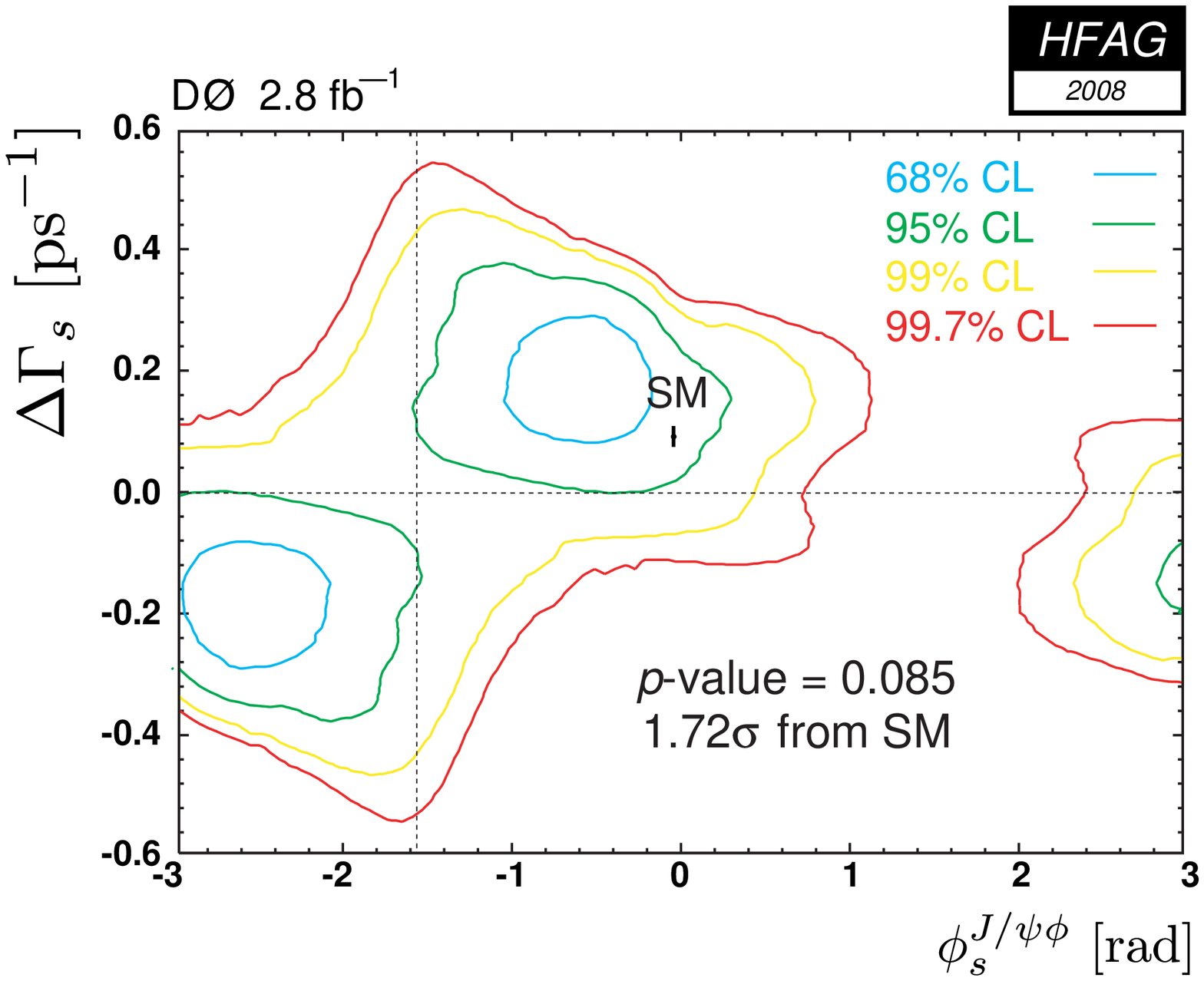}
\includegraphics[width=7.0cm]{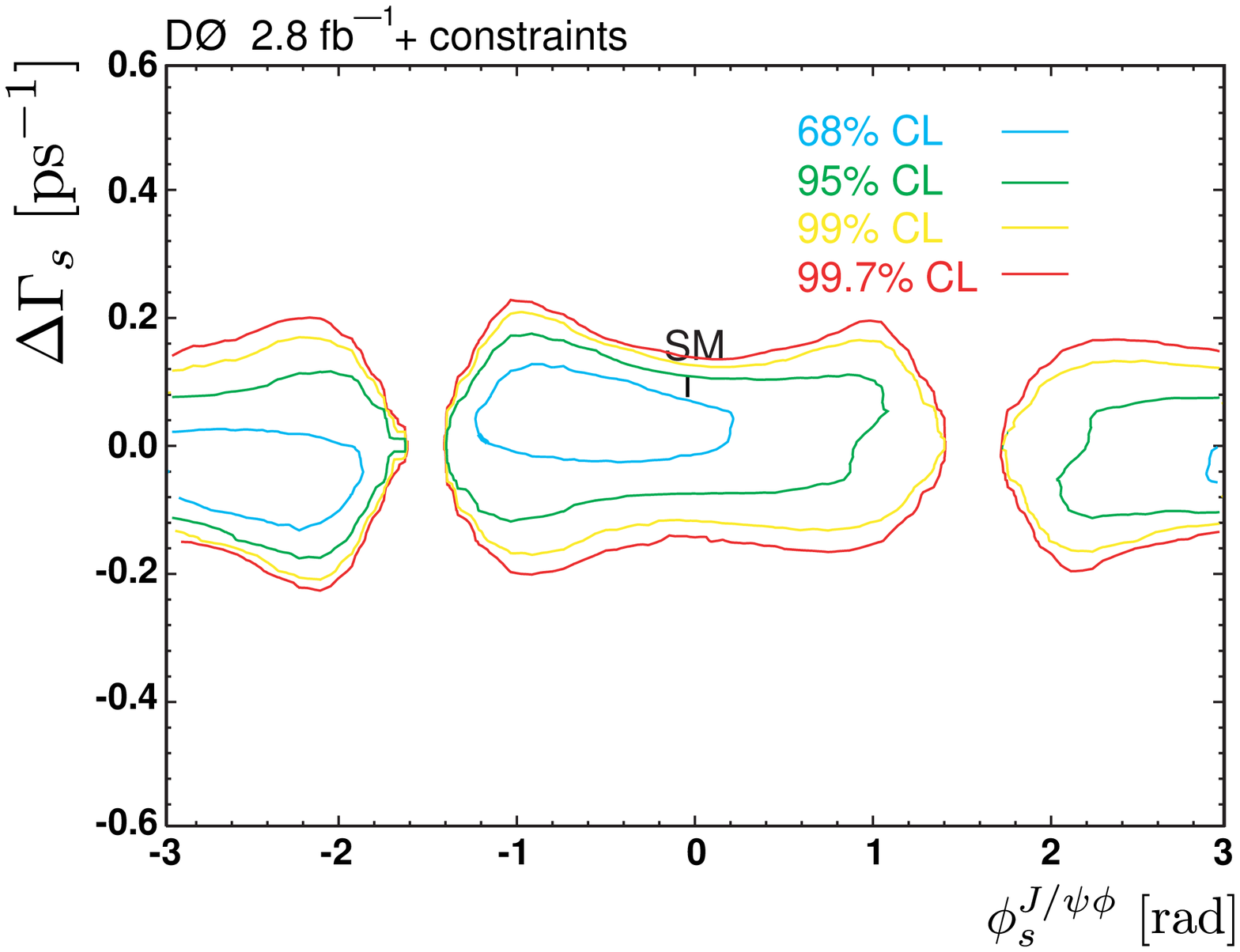}
\unitlength 1cm
\begin{picture}(0.1, 0.1)
  \put(10.5,1.9){\sf \scriptsize \scalebox{1.15}{$1.65\sigma$ from SM}}
\end{picture}
\vspace*{-3ex}
\caption{\label{HFAG_phi_s}
Two dimensional contour plots of the parameter fit in the ($\Delta\Gamma_s, \phi_s$) plane
for different confidence levels. The left plot does not contain any constraints.
The right plot takes constraints of the measured charge asymmetry $A^s_{SL}$ and $B^0_s$ lifetime 
into account.
}
\end{figure}

\begin{figure}[p]
\hspace*{-3ex}
\includegraphics[width=7.1cm]{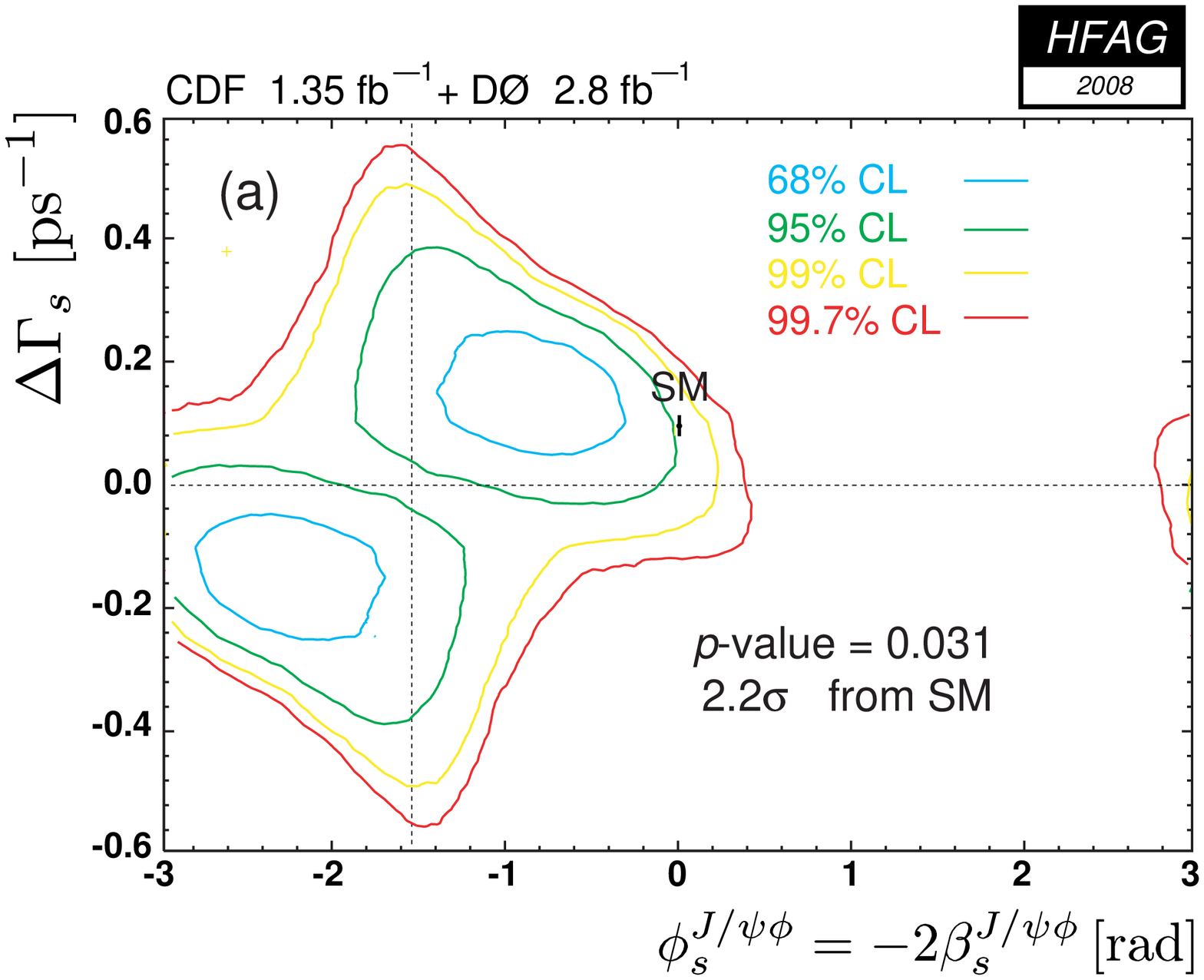}
\includegraphics[width=7.0cm]{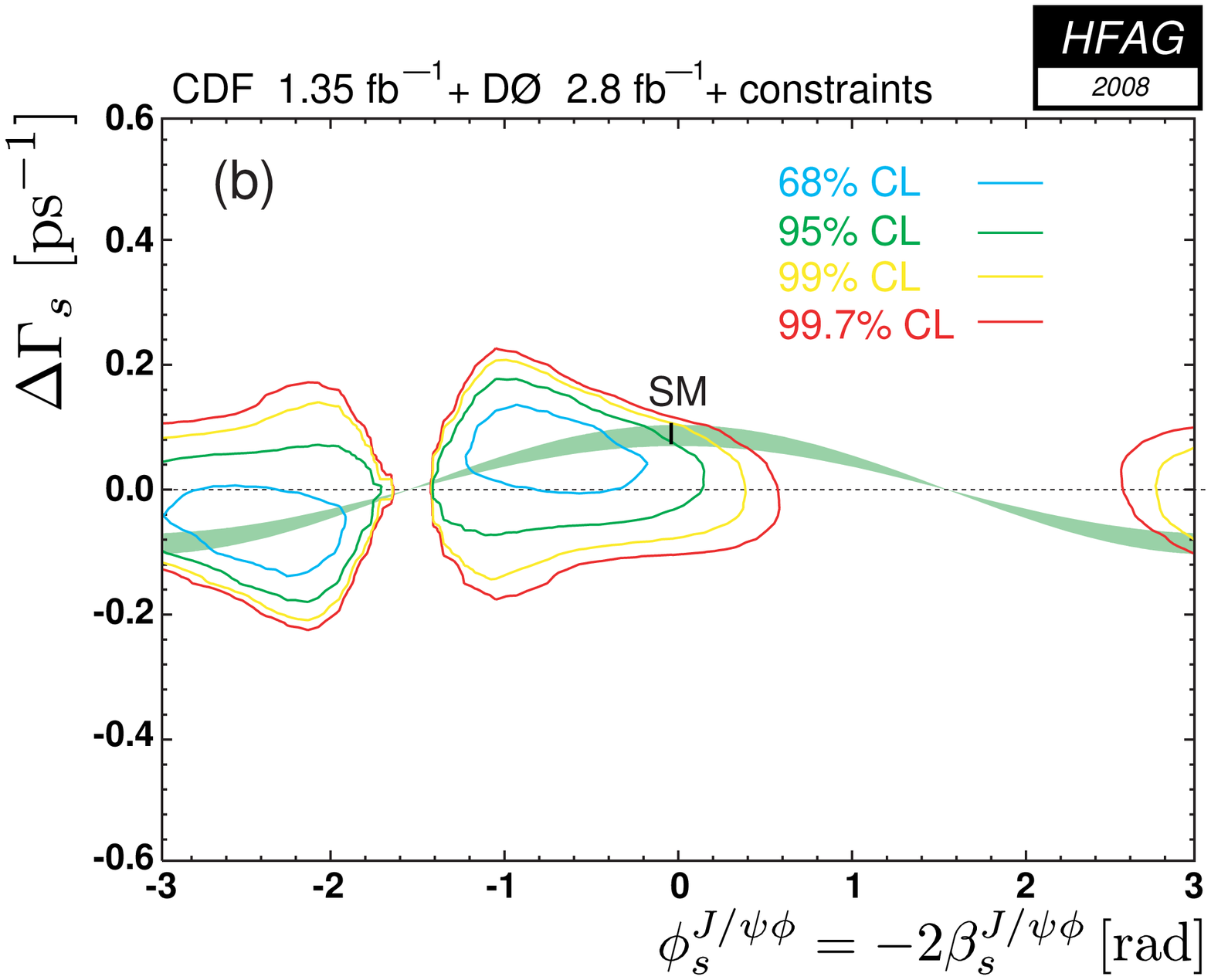}
\unitlength 1cm
\begin{picture}(0.1, 0.1)
  \put(10.9,2.1){\sf \scriptsize \scalebox{1.1}{$2.3\sigma$ from SM}}
\end{picture}
\vspace*{-3ex}
\caption{\label{HFAG_DO_CDF_phi_s}
Contour plots of D\O\ and CDF combined results in the ($\Delta\Gamma_s, \phi_s$) plane
for different confidence levels. The left plot does not contain any constraints.
The right plot takes constraints of the measured charge asymmetry $A^s_{SL}$ and $B^0_s$ lifetime into account.
}
\end{figure}

$B^0_s$ mixing parameters~\cite{f1}\cite{f2} have been measured~\cite{p3} by means of the flavor tagged decay 
$B^0_s\rightarrow J/\!\psi\phi$, exploiting an integrated luminosity of 2.8~fb$^{-1}$.
The decay width difference between the light and heavy mass eigenstates is determined to
\begin{equation}
 \Delta\Gamma_s\equiv (\Gamma_L-\Gamma_H) =  0.19\pm 0.07~(\mbox{stat}) ^{+0.02}_{-0.01}~(\mbox{syst})~\mbox{ps}^{-1}
\end{equation}
and the CP violation phase to
\begin{equation}
  \phi_s= -0.57^{+0.24}_{-0.30}~(\mbox{stat}) ^{+0.07}_{-0.02}~(\mbox{syst})~\mbox{rad} ,
\end{equation}
taking constraints on strong phases into account.
The results correspond to a $1.8\sigma$ deviation from the SM.
Fig. \ref{phi_s_contour} shows the two dimensional contour plot of the measurement in the ($\Delta\Gamma_s, \phi_s$) 
plane.
In Fig. \ref{HFAG_phi_s} the same kind of contour plots 
%from the Heavy Flavor Averaging Group (HFAG~\cite{h2}),
are given, without any constraints (left plot) and taking into account constraints from the measured charge asymmetry $A_{SL}^s$  and $B^0_s$ lifetime (right plot).
%are given.
The preliminary time dependent tagged D\O\ analysis mentioned above, measuring the CP violation parameter $a^s_{sl}$ is not included yet. The dataset is assumed to be tripled until end of 2009 and a 40\% increase 
in reconstruction efficiency is anticipated leading to an projected increase in significance of about a factor of two.
The combinations of present D\O\ and CDF results~\cite{c2}\cite{c3} 
from the Heavy Flavor Averaging Group~\cite{h2} are presented in Fig. \ref{HFAG_DO_CDF_phi_s} without any constraints 
(left plot) and with constraints from charge asymmetry $A^s_{SL}$
and $B^0_s$ lifetime measurements 
(right plot).

\section*{Summary}
The CP violation phase $\phi_s$ and charge asymmetries in $B^0_s$ decays have been measured
with the D\O\ detector. 
A $1.8\sigma$ deviation from the Standard Model is observed in the measurement of the
CP violation phase $\phi_s= -0.57^{+0.24}_{-0.30}$~(stat) $^{+0.07}_{-0.02}$~(syst) rad
in the decay channel $B^0_s\rightarrow J/\!\psi\phi$, taking constraints on strong phases into account.
Without any constraints the deviation changes marginally to $1.72\sigma$.
Taking into account constraints of charge asymmetry $A^s_{SL}$ and $B^0_s$ lifetime measurements 
this value alters to $1.65\sigma$.
A combination of the D\O\ results with the measurement of CDF
yields a $2.2\sigma$ deviation from the SM if no constraints are applied.
Constraints from charge asymmetry $A^s_{SL}$ and $B^0_s$ lifetime measurements
increase the deviation slightly to $2.3\sigma$.

\section*{Acknowledgements}
Many thanks to the stuff members at Fermilab, collaborating institutions and the $B$ physics group
of the D\O\ experiment.
Among others, this work has been supported by the Marie Curie Program,
in part under contract number MRTN-CT-2006-035606
and the HEPtools EU Marie Curie Research Training Network under contract number
MRTN-CT-2006-035505.

\end{document}